\documentclass{svjour3}                     
\smartqed  
\usepackage{graphicx}
\usepackage{mathptmx}      
\journalname{Journal of Low Temperature Physics}
\begin{document}

\title{Shedding of Vortex Rings from an Oscillating Sphere in Superfluid $^4$He below 0.5 K -- The Origin of the Turbulent Drag Force}


\titlerunning{Oscillating Sphere}        

\author{W. Schoepe        
}


\institute{W. Schoepe \at
              Fakult\"at f\"ur Physik, Universit\"at Regensburg, Germany \\
              \email{wilfried.schoepe@physik.uni-regensburg.de}           
}

\date{}

\maketitle

\begin{abstract}

The onset of turbulent flow around an oscillating sphere is known to occur at a critical velocity $v_c \sim \sqrt{\kappa \,\omega }$ where $\kappa $ is the circulation quantum and $\omega $ is the oscillation frequency. However, in a small interval of driving force amplitudes $F$ (or corresponding velocity amplitudes of few percent above $v_c$) the turbulent flow is found to be unstable. The flow pattern switches intermittently between potential flow and turbulence. The lifetimes of the turbulent phases have an exponential distribution and the mean lifetimes $\tau $ grow very rapidly with increasing driving force, namely  as $\tau (F) \propto \exp[(F/F_1)^2]$. In this work this experimental result is analyzed in more detail than before, in particular the force $F_1$ is identified. As a result, the turbulent drag force $F(v) \propto (v^2 - v_c^2)$ can be ascribed quantitatively to the shedding of vortex rings having the size of the sphere. Moreover, we can infer the average number of vortex rings that are shed per half-period at any given velocity $v$ on the turbulent drag force.

\end{abstract}

\section{Introduction}
\label{intro}
Laminar and turbulent flow of superfluid $^4$He around an oscillating microsphere (radius $R$ = 0.12 mm) has been studied in detail since 1995 \cite{PRL1,OVL}. At small oscillation amplitudes the drag force is linear in velocity amplitude (Stokes regime) and is determined by the normal fluid component. At temperatures below ca. 0.5 K the linear drag is given solely by ballistic phonon scattering and vanishes as $T^4$. Above a critical velocity amplitude \,$v_c = 2.8 \sqrt{\kappa \omega }$\, \cite{arx,risto1} a transition to a large and nonlinear drag force is observed that scales as $\gamma \,(v^2 - v_c^2)$, see Fig.1. (Here we are using the terms ``drag force" and ``driving force" synonymously. Strictly speaking, for our quadratic drag force a ca. 15\% correction due to the principle of energy balance is needed  \cite{JLTP150}.) From a fit to the data we find that $\gamma $ is given by the classical value \cite{LL} $c_D \rho \pi R^2/2$, where $\rho$ is the density of the superfluid and the coefficient $c_D$ = 0.36 is in agreement with what is known from spheres in fully developed classical turbulence. It is remarkable that for oscillating turbulent superflow we find $\gamma$\, to be identical to the value for classical turbulent dc flow. The shift of the drag force by $\gamma \, v_c^2$ is the signature of the superfluid: At $T$ = 0 an infinitesimal driving force will lead to the critical velocity $v_c$. This transition is obviously the onset of turbulent flow. For velocities within a few percent above $v_c$ an instability of the flow is observed at temperatures below ca. 0.5 K, namely an intermittent switching of the flow pattern between turbulent and laminar phases \cite{OVL}, see Fig.2 where three time series are shown that were recorded at a constant temperature of 300 mK and at different driving forces. During a turbulent phase the drag is given by the large turbulent drag force and, hence, the velocity amplitude is low. When turbulence breaks down the velocity amplitude begins to grow due to the much smaller phonon drag. Because the velocity amplitudes are above $v_c$ these phases of laminar flow (better: potential flow) break down and a new turbulent phase is observed. It is obvious that the lifetimes of the turbulent phases grow very rapidly with the driving force. In the following, the lifetimes of these phases of turbulent flow will be discussed in more detail.\\

\begin{figure}
\begin{center}
\includegraphics[width=0.9\linewidth, keepaspectratio]{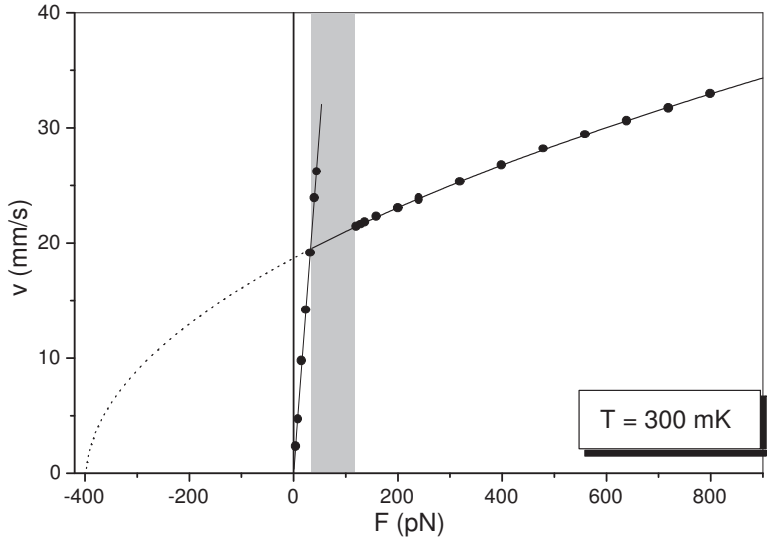}
\end{center}
\caption{(From Ref.2) Velocity amplitude $v$ of the oscillating sphere (frequency 114 Hz) as a function of the driving 
force $F$ at $300\,\mbox{mK}$. At small drives the flow is laminar and the linear 
behavior is determined by ballistic phonon scattering. The shaded area above the critical velocity $v_c \approx $ 20 mm/s indicates the unstable regime where the flow switches intermittently between potential flow and turbulence. At larger driving forces we observe apparently stable nonlinear turbulent drag where $F(v)$ scales as $\gamma (v^2 - v_c^2)$ (when the small laminar drag is subtracted).}

\label{fig:1}       
\end{figure}

\begin{figure*}
\begin{center} 
 \includegraphics[width=0.8\textwidth]{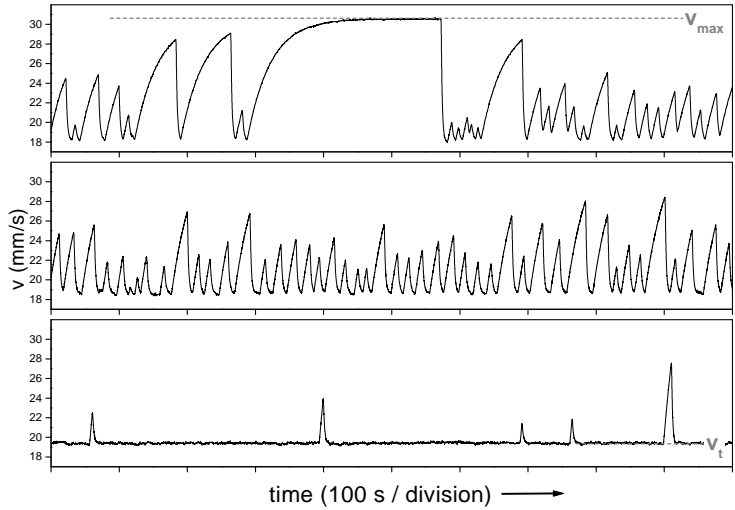}
\end{center}
\caption{(From Ref.2) Three time series of the velocity amplitude at 300 mK and at three different driving forces (in pN): 47, 55, 75 (from top to bottom). The low level $v_t$ corresponds to turbulent flow while the increase of the velocity amplitude occurs during a laminar phase, occasionally reaching the maximum value $v_{max}$ given by the linear drag. With increasing drive the lifetimes of the turbulent phases grow rapidly.}
\label{fig:2}       
\end{figure*}

\section{Lifetimes of Turbulent Phases}

\begin{figure}
\begin{center}
\includegraphics[width=1.0\textwidth]{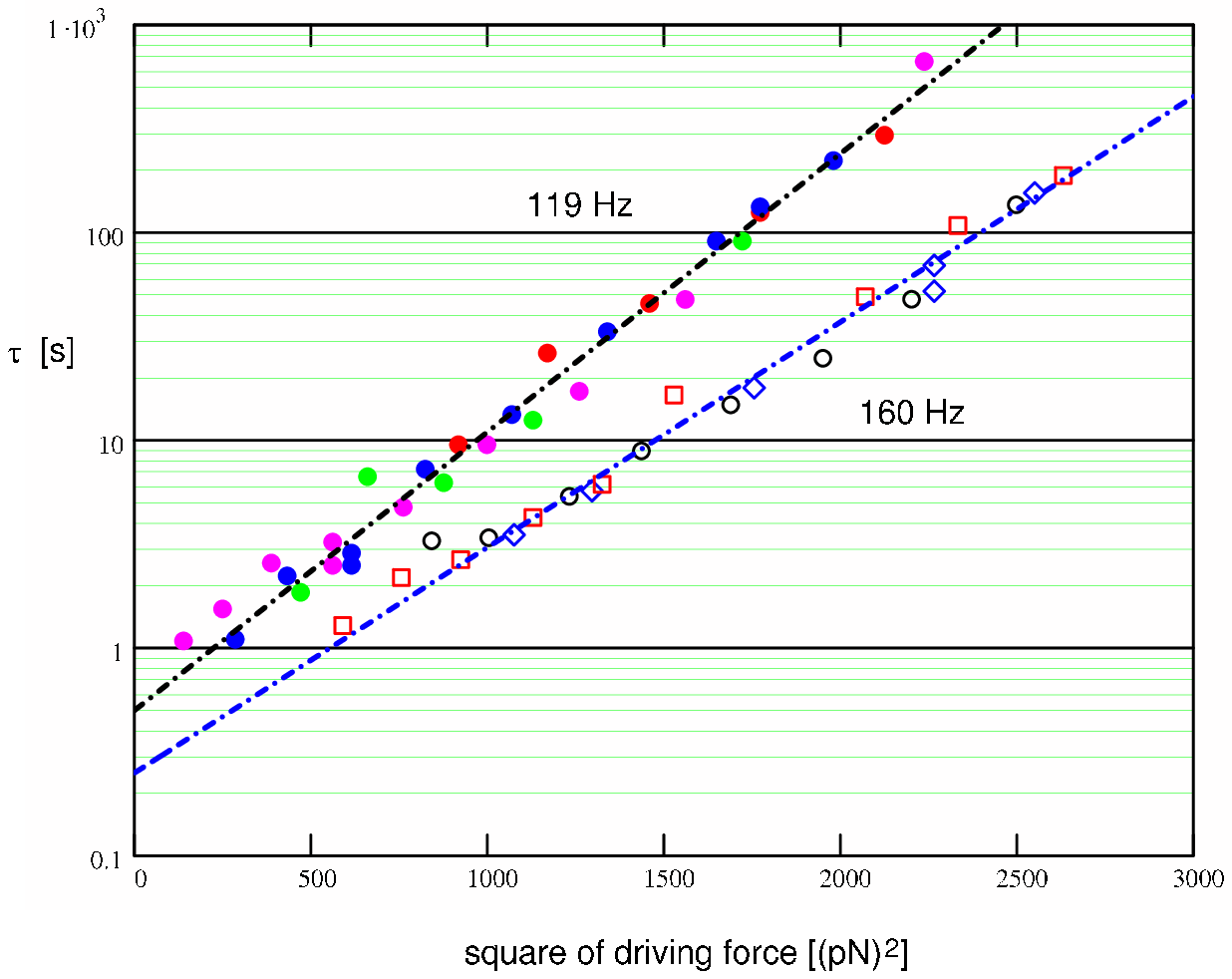}
\end{center}
\caption{(Color online) Mean turbulent lifetimes as a function of the driving force at two different oscillator frequencies. Each data point is obtained from a time series, some of which lasted up to 36 hours.The straight lines are fits of (1) to the data. The data of the 119 Hz oscillator were taken at 4 different temperatures (in mK): red 403, blue 301, green 200, and violet 100. The data at 160 Hz are: black circles 300 mK, red squares 30 mK, mixture with 0.05\% $^3$He, blue diamonds 30 mK with 0.5\% $^3$He. Note that the slopes $1/F_1^2$ and the intercepts $\tau _0$ are independent of temperature and $^3$He concentration, but both depend on the oscillation frequency.}

\label{fig:3}       
\end{figure}

The turbulent phases are found to be exponentially distributed as $\exp(-\,t /\tau)$ \cite{OVL}, and the mean lifetimes $\tau $ are found to increase very rapidly with increasing driving force $F$, namely as 
\begin{equation}
\tau (F) = \tau_0 \,\, \exp[\,(F/F_1)^2\,], 
\end{equation}
see Fig.3, until they become longer than the observation times, and, hence, turbulence {\it appears} to be stable. In our earlier work, this switching behavior has been discussed in some detail \cite{OVL,PRL2,Lammi2010}. In this work, a new and more detailed and more quantitative analysis of the mean turbulent lifetimes is presented. In particular, the ``characteristic" force $F_1$ is identified: it will be shown that it is due to the shedding of vortex rings of the same size $R$ as the oscillating sphere. This is a very plausible result, of course, but - to our knowledge - has not been derived quantitatively before. Instabilities of the flow and switching phenomena due to emission of vortex rings have been reported with vibrating wires in superfluid $^4$He \cite{Kubo} and $^3$He \cite{Bradley} but no quantitative analysis of the turbulent drag force has been possible. Also, it should be mentioned that vortex shedding from a spherical pendulum in water has been observed and beautiful pictures of vortex rings are presented \cite{Russ2}. Because of the viscosity of water and, hence, because of the existence of a viscous boundary layer on the sphere the analysis of the freely decaying pendulum oscillations is quite different from our analysis of the transient turbulence.\\ 

Our results for the mean turbulent lifetimes are displayed in Fig.3.
We obtain for the 119 Hz oscillator $\tau _0$ = 0.5 s and $F_1$ = 18 pN, while for the 160 Hz oscillator we have $\tau _0$ = 0.25 s and $F_1$ = 20 pN. Although we have data only for two frequencies \footnote{The resonance frequency of the sphere was changed by warming the cell briefly above 10 K, for experimental details, see \cite{PRL1,OVL}.} it appears that $F_1$ scales as $\sqrt {\omega }$ because $\sqrt {160/119}$ = 1.16 and 20/18 = 1.11. In the following we will first concentrate on the properties of $F_1$ by using dimensional arguments.

\section{Dimensional Considerations}
In addition to its frequency dependence the force $F_1$ must be given by a product of the density of liquid helium $\rho = 145$ kg/m$^3$, a length scale which we take as the radius $R = 0.12 $ mm of the sphere, and the circulation quantum $\kappa \approx 10^{-7}$ m$^2$/s.\\
\\
The result is
$F_1 \propto \rho \,\kappa ^{3/2} \,R \,\sqrt{\omega} = \rho \,\kappa \,R\, \sqrt{\kappa\, \omega}$.
The latter expression is more appropriate because $\sqrt{\kappa\, \omega}$\, determines the critical velocity, see above.
Of course, we must allow for a numerical factor $c$ which can be obtained only from the experiment:

\begin{equation}  
F_1  = c\, \rho \, \kappa \,R \,\sqrt{\kappa \,\omega}.
\end{equation}
From our experimental results for $F_1$ we find $c \approx 1.3$. What is the origin of $F_1$?

\section{A Model for the ``Characteristic" Force \boldmath $F_1$}

We propose that a vortex ring of the size $R$ is shed by the sphere during each half-period $T/2 = \pi /\omega $. The power lost by the sphere amounts to the energy $E(R)$ of the ring per half-period. For $E(R)$ we take the usual formula, namely \cite{Russ}
\begin{equation} 
E(R) = 1/2\,\, \rho \kappa ^2 \, R [\,\ln( 8R/a_0) - 7/4].
\end{equation}
The term in the brackets is roughly 14.
The dissipated power is then given by\\ 
\begin{center}
$\frac{1}{2} \, F_1 v_c = E(R) \omega /\pi $,\\
\end{center}
i.e.,
\begin{equation}
F_1 = (14/2.8 \,\pi )\, \rho\,\kappa \,R\, \sqrt{\kappa \,\omega} = 1.6 \, \rho\,\kappa \,R\, \sqrt{\kappa \, \omega}.\\
\end{equation}
This result agrees with (2) and with our experiment: $F_1$ scales as $\sqrt{\omega }$ and is independent of temperature, the numerical factor of 1.6 is close to our experimental value of 1.3. We interpret the number $n = F/F_1$as the average number of vortex rings emitted per half-period. Our data in Fig.3 lie in the range $0.7 \le n < 3$.

\section{Changing Variables: from Forces to Velocities}

The drag force determines the velocity amplitude of the sphere: $F(v) = \gamma \,\,(v^2 - v_c^2)$. We write
\begin{equation}
n = \frac{F}{F_1} = \frac{\gamma \,\,(v^2 - v_c^2)}{1.3\,\,\rho \,\kappa \,R\, \sqrt{\kappa \,\omega}} = \frac{\Delta v}{v_1}(1+ \frac{\Delta v}{2 v_c}),
\end{equation}
where $\Delta v = v - v_c$ and the ``characteristic" velocity is $v_1 = 0.41 \,\,\kappa / R$ = 0.34 mm/s, and is independent of any other quantity. We note that $\kappa /R$ is of the same order of magnitude as the self-induced velocity of a vortex ring of size $R$ \cite{Russ}.\\
We can assign now to any velocity coordinate of $F(v)$ both the number $n$ of vortex rings that are shed per half-period and the lifetime $\tau = \tau _0 \,\exp( n^2 )$. For example, for $\Delta v/v_c$ = 0.05 we find $n$ = 3.65 and $\tau = 3 \cdot 10^5$ s (for the 119 Hz oscillator), while for $\Delta v/v_c$ = 0.1 we have $n$ = 7.48 and $\tau \sim 10^{24}$ s, which is much longer than the lifetime of the Universe.
From (5) also follows that the drag force can now be written as $F(n) = 2\, n \,\gamma \,v_1\,v_c$.\\

For very small values of $\Delta v/v_c \le 0.03$, i.e., in the regime where we observe the intermittent switching, we may neglect the term $ \Delta v/2 v_c$ in (5) and simply use $n = \Delta v /v_1$, i.e., we can plot the normalized lifetimes
\begin{equation}  
\tau^* (\Delta v) = \tau / \tau _0 = \: \exp[(\Delta v/v_1)^2], 
\end{equation}
see Fig.4. 
Note the very rapid growth of $\tau^* $ in a very small interval of velocities ($ \approx 2\%$ above $v_c$) and the frequency independence of the data.\\

From Fig.3 the prefactors $\tau _0$ cannot be determined as accurately as the forces $F_1$. Moreover, since we have data only for two oscillation frequencies, the frequency dependence of $\tau _0$ remains unclear at present, see below.

\begin{figure}
\begin{center}
\includegraphics[width=0.7\textwidth]{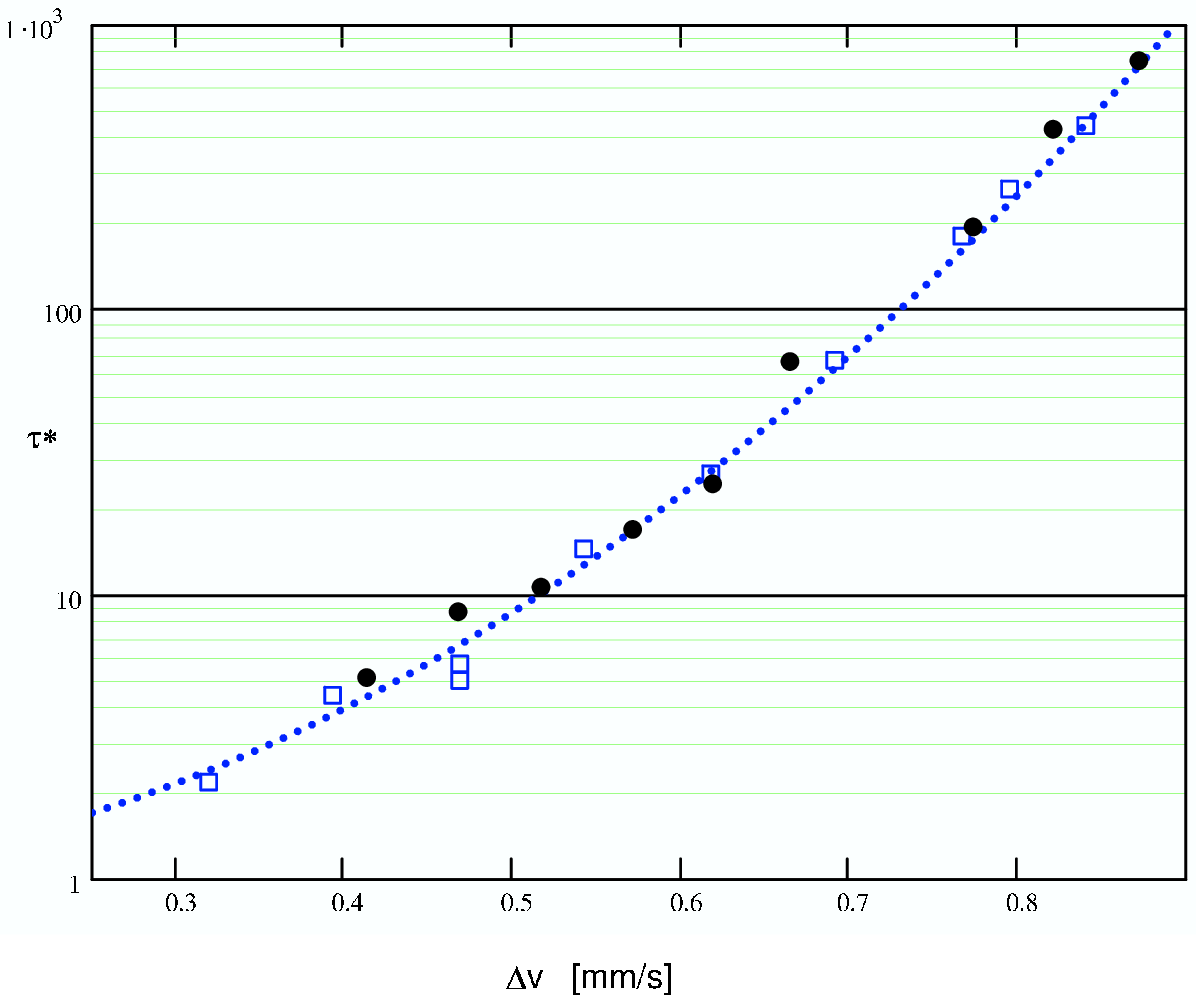}
\end{center}
\caption{(Color online) The normalized lifetimes $\tau ^* = \tau /\tau_0$ as a function of $\Delta v = v - v_c$ for the 119 Hz oscillator at 301 mK (blue squares) and the 160 Hz oscillator at 30 mK with 0.05\% $^3$He (black dots). Note the rapid increase of $\tau ^*$ by almost 3 orders of magnitude over the small velocity interval of ca. 0.5 mm/s and the frequency independence of the data.} 
\label{fig:4}       
\end{figure}

\section{The Rice Formula}
The exponential dependence of $\tau $ on $(\Delta v / v_1)^2$ may be described by Rice's formula \cite{Rice,Rootzen} for Gaussian fluctuations crossing a given level. It is applied here to fluctuations of $v$ crossing $v_c$ where turbulence breaks down. The average number per unit time of crossings of a given level $C$ below a mean at zero with negative slope is proportional to $\exp(- C^2/2\sigma ^2)$ with $\sigma ^2$ being the variance of the fluctuations. Postulating that the lifetime of a turbulent phase is ending at the first crossing of $v$ of level $v_c$ \cite{PRL2}, we obtain (6). $\tau _0 $ is interpreted to be twice the average time between two crossings of the level $v_c$ \cite{Rice}, i.e., for $\tau _0$ = 0.5 s or 0.25 s there are 4 or 8 crossings per second which means that the velocity of the sphere fluctuates at an average frequency of 2 Hz or 4 Hz, respectively. This could not be detected with our electronics. 


\section{Summary}

Based on a quantitative analysis of the measured lifetimes of the turbulent flow around the oscillating sphere we have achieved a more detailed understanding of what causes the turbulent drag on the sphere. Our model of vortex shedding from the sphere is plausible and describes quantitatively our experimental results on the transition to quantum turbulence. In our earlier work \cite{PRL2,Lammi2010} we had inferred the vortex line density from the velocity amplitude by making use of results based on dc vortex dynamics. As long as a theory of oscillating turbulent superflow is not available, these assumptions are not yet proven to be valid. In contrast, in our present work we are discussing only the forces $F$ and the resulting velocity amplitudes $v(F)$ and, therefore, avoiding those uncertainties.\\
\\
Finally, we emphasize again (as done already earlier \cite{JLTP150}): when the rapidly growing lifetimes of the turbulent phases begin to exceed the time of observation, driven superfluid turbulence {\it appears to be stable}, which is misleading because the lifetimes are {\it large but finite and do not diverge}. A similar result has been found in turbulent classical pipe flow \cite{Hof}.

\begin{acknowledgements}
A discussion with my former co-workers Hubert Kerscher and Michael Niemetz as well as comments from Matti Krusius (Aalto University, Helsinki) are gratefully acknowledged. 
\end{acknowledgements}

\end{document}